\pdfoutput=1
\documentclass[nofootinbib,aps,11pt]{revtex4}
\usepackage{graphicx}
\usepackage{cancel}
\usepackage{amssymb}
\usepackage{textcomp}
\usepackage{amsmath}
\usepackage{bm}
\usepackage{times}
\usepackage{epsfig}
\usepackage{color}
\def\ie{{\it i.e.}}

\definecolor{evgray}{gray}{.48}
\definecolor{orange}{RGB}{255,127,0}
\definecolor{purple}{RGB}{147,112,219}
\definecolor{dgreen}{RGB}{0,180,10}

\newcommand{\espaBig}{\vphantom{\displaystyle{\int_{\frac{a}{b}}^{\frac{a}{b}}}}}
\newcommand{\xmapsto}[1]{\xrightarrow{\ #1\ }}

\graphicspath{{./pictures/}}

\begin{document}

\title{\Large
Constraints to Dark Matter from Inert Higgs Doublet Model
}

\author{Marco Aurelio D\'{\i}az, Benjamin Koch, and Sebasti\'an Urrutia-Quiroga$^\ast$ \\[2mm]
{\small
{\it Instituto de F\'\i sica, Pontificia Universidad Cat\'olica de Chile, Av. Vicu\~na Mackenna 4860, Santiago, Chile}}
\\
{\small $^\ast$\tt{sgurruti@uc.cl}}
}

\begin{abstract}

We study the Inert Higgs Doublet Model and its inert scalar Higgs $H$ as the only source for
dark matter. It is found that three mass regions of the inert scalar Higgs can give the correct dark matter relic density.
The low mass region (between 3 and 50 GeV) is ruled out.
New direct dark matter detection experiments will probe the intermediate (between 60 and 100 GeV) and high
(heavier than 550 GeV) mass regions. Collider experiments are advised to search for $D^\pm\rightarrow HW^\pm$
decay in the two jets plus missing energy channel.

\end{abstract}

\maketitle

\section{Introduction}
\label{secIntro}

Astrophysical observations provide strong evidence for the existence of Dark Matter (DM) \cite{Clowe:2006eq},
and its abundance in the current phase of the Universe \cite{Ade:2013sjv}.
According to the newest results from Planck collaboration,
there is a 74\% approximately of matter which is not directly visible but
is observed due to its gravitational effects on visible matter.
Additional evidence for the existence of DM comes from study the rotation
curves of spiral galaxies \cite{Rubin:1980zd}, the analysis of the bullet cluster \cite{Markevitch:2003at},
and the study of baryon acoustic oscillations \cite{Percival:2007yw}.
One of the most common hypothesis used to explain these phenomena is to postulate
the existence of weakly interacting massive particles (WIMPs) \cite{Copi:1994ev}.

In order to explain DM, we study a simple extension of the Standard Model (SM), called the Inert Higgs Doublet Model (IHDM).
This model, which was originally proposed for studies on electroweak (EW) symmetry breaking \cite{Deshpande:1977rw}, introduces an additional doublet and a discrete symmetry.
These two characteristics of the model modify the SM phenomenology, but there are some regions of the parameter space which predict only
small deviations from the SM.
Nevertheless, one of the most attractive characteristic of the IHDM is the presence of an stable neutral particle which can be a DM candidate.

In this work the IHDM is revisited, considering various restrictions,
focusing our analysis on the zones of the parameter space which reproduce the correct DM relic density according to the newest measurements \cite{Blinov:2015qva, Khan:2015ipa}.
In this context, three not connected mass regimes for the lightest inert particle are found.
These regimes are also analyzed using LHC observables like branching ratios to invisible particles and an specific SM--like Higgs boson decay mode.
Additionally, we study some inert decays modes. Finally, we use a direct detection approach to rule out one of the mass regimes.
It is further shown that this regime can also be ruled out by constraints from collider physics \cite{Belanger:2015kga}.

This paper is organized as follows.
After a short introduction in section \ref{secIntro}
 the model is introduced in section~\ref{secHDM} by formulating the associated potential and constraints of the model,
 and by exploring the parameter space and its characteristics in section \ref{secParameters}.
 In section \ref{secCollider} the behavior of the model is presented from a collider physics perspective, studying the modifications of the SM and its implications for the IHDM. In section \ref{secDM} the results of this study are complemented
 by an analysis from the Dark Matter perspective.
 Finally, in section \ref{secConc} we remark on the most important conclusions of our work.

\section{The Inert Higgs Doublet Model}
\label{secHDM}

Consider an extension of Standard Model (SM), which contains two Higgs doublets $\Phi_{S,D}$ and a
 discrete $\mathbb{Z}_2$ symmetry \cite{Deshpande:1977rw}  ($\Phi_S \xmapsto{\mathbb{Z}_2} \Phi_S$ and
$\Phi_D \xmapsto{\mathbb{Z}_2} -\Phi_D$). All fields of the SM are invariants under the discrete symmetry,
and $\Phi_S$ is completely analogous to the SM Higgs doublet.

The most general renormalizable $SU(2) \times U(1)$ invariant Higgs potential, that also preserves
the discrete symmetry is,
\begin{equation}
V=\mu_1^2\boldsymbol{A}+\mu_2^2\boldsymbol{B}+\lambda_1\boldsymbol{A}^2+\lambda_2\boldsymbol{B}^2+\lambda_3\boldsymbol{A}\boldsymbol{B}+\lambda_4\boldsymbol{C}^\dagger \boldsymbol{C}+\frac{\lambda_5}{2}\Big(\boldsymbol{C}^2+\boldsymbol{C}^{\dagger 2}\Big)\quad,
\label{V}
\end{equation}
where $\boldsymbol{A},\boldsymbol{B},\boldsymbol{C}$ are given by,
\begin{equation}
\boldsymbol{A}={\Phi_S}^\dagger\Phi_S,\,\boldsymbol{B}={\Phi_D}^\dagger\Phi_D,\,\boldsymbol{C}={\Phi_S}^\dagger\Phi_D\quad.
\end{equation}
The parameters $\mu_i^2$ $(i=1,2)$ and $\lambda_j$ $(j=1\ldots4)$ are intrinsically real,
and $\lambda_5$ will be assumed to be real \cite{Ginzburg:2010wa}. After Spontaneous Symmetry Breaking, the vacuum expectation values of the
Higgs doublets are,
\begin{equation}
\langle\Phi_S\rangle=\frac{1}{\sqrt{2}}\left(\begin{array}{c}0\\v\end{array}\right),\quad\langle\Phi_D\rangle=\left(\begin{array}{c}0\\0\end{array}\right)\quad,
\end{equation}
where $\langle\Phi_D\rangle$ is forced by the discrete symmetry, and $v=246\,\textrm{GeV}$.
Expanding the fields around those vacua,
we define,
\begin{equation}
\Phi_S=\left(\begin{array}{c}G^+\\(v+h+iG^0)/\sqrt{2}\end{array}\right),\quad\Phi_D=\left(\begin{array}{c}D^+\\(H+iA)/\sqrt{2}\end{array}\right)\quad,
\label{particles}
\end{equation}
where $G^0$ and $G^\pm$ are the neutral and charged Goldstone bosons, and $h$ is the SM-like Higgs boson.
The fields in the second doublet belong to the so called {\it ``dark''} or {\it ``inert''} sector. They
are the scalar $H$ and pseudoscalar $A$, both neutral, and the charged scalar $D^\pm$. As in the SM, the
parameter $\mu_1^2$ is related to $\lambda_1$ by the tree-level tadpole condition $\mu_1^2=-\lambda_1v^2$.
The masses of physical states \cite{Honorez:2010re, Arhrib:2013ela} are,
\begin{eqnarray}
m_h^2 &=& 2\lambda_1 v^2 \quad,\cr
m_D^2 &=& \mu_2^2+\frac{\lambda_3}{2}v^2 \quad,\cr
m_H^2 &=& m_D^2+\left(\frac{\lambda_4+\lambda_5}{2}\right)v^2 = \mu_2^2+\dfrac{\lambda_{345}}{2}v^2 \quad, \cr
m_A^2 &=& m_D^2+\left(\frac{\lambda_4-\lambda_5}{2}\right)v^2\quad,
\label{masses}
\end{eqnarray}
with
\begin{equation}
\lambda_{345} = \lambda_3+\lambda_4+\lambda_5\quad.
\label{lambdaT}
\end{equation}
As independent and free parameters we take the masses $m_H$, $m_A$, and $m_D$ at tree-level, and the couplings
$\lambda_2$ and $\lambda_{345}$. The SM-like Higgs boson mass is fixed now thanks to the measurement
$m_h=125$ GeV \cite{:2012gk,Chatrchyan:2013lba}.

The constraints we initially impose includes vacuum stability at tree-level,
where constraints on the $\lambda_i$ couplings and $\mu_i$
mass terms appear \cite{Gustafsson:2010zz, Fortes:2014dia, Krawczyk:2013pea};
perturbativity ($|\lambda_i|<8 \pi$) \cite{Barbieri:2006dq, Djouadi:2005gi} and unitarity \cite{Arhrib:2012ia},
where we impose that the scalar potential is unitary and that several scattering processes between scalar
and gauge bosons are bounded; electroweak precision tests through the $S$, $T$, and $U$ parameters
\cite{Peskin:1991sw} applied to the IHDM \cite{Arhrib:2012ia, Swiezewska:2012ej}, with $3\sigma$ values
given by
\begin{equation}
S\Big|_{U=0} = 0.06 \pm 0.09 \qquad\wedge\qquad T\Big|_{U=0} = 0.10 \pm 0.07 \quad,
\end{equation}
with a correlation coefficient of $+0.91$ \cite{Baak:2014ora} and collider constraints
\cite{Gustafsson:2010zz, Lundstrom:2008ai, Cao:2007rm, Pierce:2007ut},
where we satisfy lower bonds on the Higgs boson masses.

The DM particle must be neutral. In our analysis we assume it is the $H$ boson, thus $m_H<m_A$ and $m_H<m_D$,
which due to (\ref{masses}) translates  to
\begin{equation}
\lambda_4+\lambda_5<0 \quad\wedge\quad \lambda_5<0\quad.
\end{equation}

We do not consider $A$ as the DM candidate because it is analogous to consider $H$ as the DM candidate defining
$\lambda^-_{345}=\lambda_3+\lambda_4-\lambda_5$ instead of $\lambda_{345}$.

\section{IHDM Parameter Space}
\label{secParameters}

We randomly scan the parameter space of the IHDM, taking into account all the constraints mentioned
in the previous section. Additionally, we compute some astrophysical properties of the model using the \verb"micrOMEGAs" software \cite{micromegas}. We consider masses satisfying 1 GeV $< m_i<$ 1 TeV, where $i=H, A, D$. In addition,
we consider cosmological measurements: the DM relic density $\Omega_{DM}\,h^2$ is a property related
to its abundance in the current phase of the Universe. This quantity is well measured by WMAP
\cite{Bennett:2012zja} and Planck \cite{Ade:2015xua} experiments. Following \cite{stat} to combine both
measurements we obtain,
\begin{equation}
\Omega_{DM}\,h^2 = 0.1181 \pm 0.0012\quad.
\label{relic}
\end{equation}
%

%
\begin{figure}[ht]
\includegraphics[height=6.5cm]{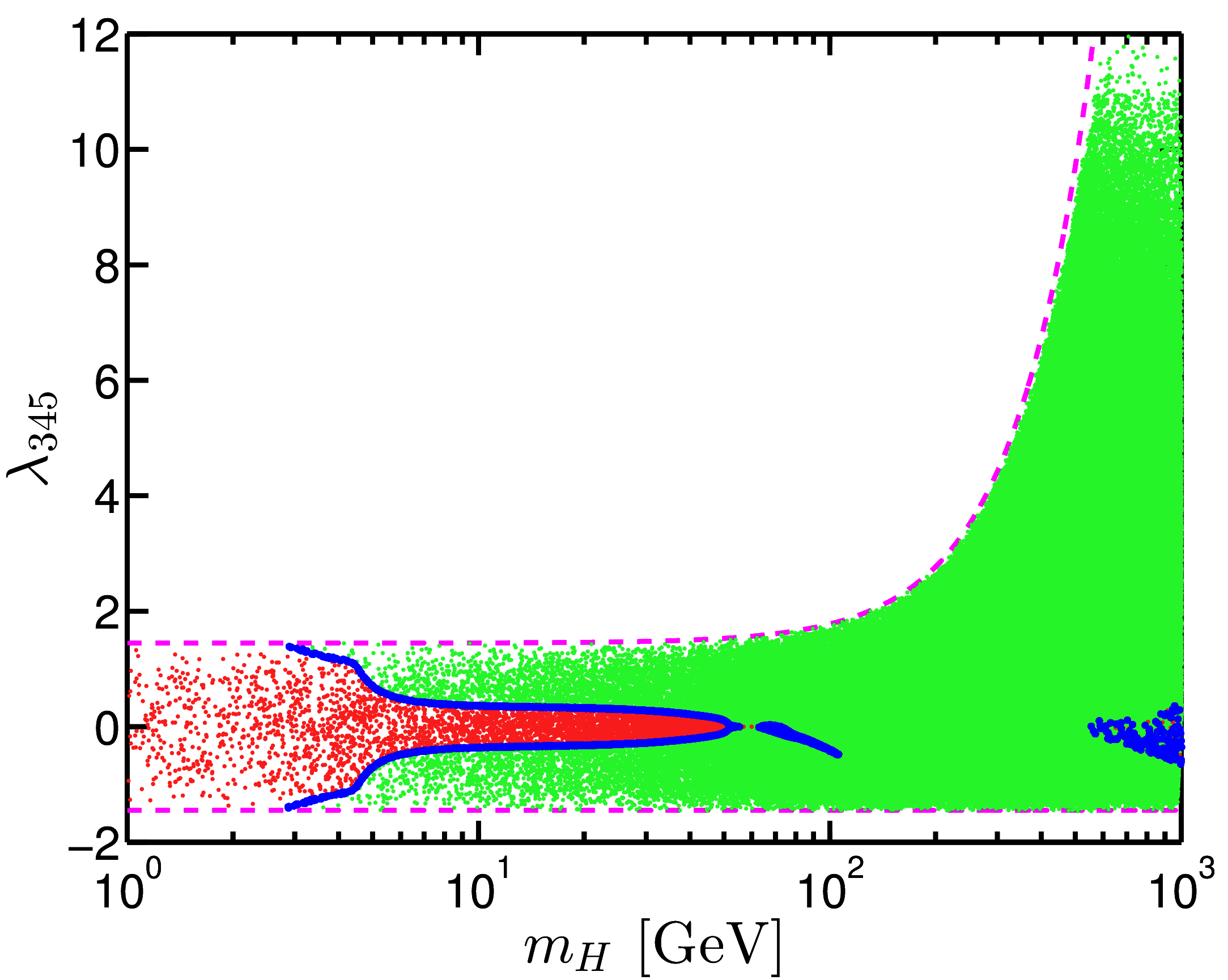}
\caption{\it Random scan of IHDM parameter space. Coupling $\lambda_{345}$ as a function of the DM candidate
mass $m_H$.
The upper dotted line is the bound generated by the inert vacuum condition and the lower dotted line
is generated by the vacuum stability condition.}
\label{FigA}
\end{figure}
%
In Figure \ref{FigA} the coupling $\lambda_{345}$, is shown as a function of the
Higgs boson mass $m_H$ varying ($\mu_i,\,\lambda_i,\,m_A,\,m_D \textrm{ and } m_H$).
We work with he hypothesis that the light inert Higgs
boson $H$ is providing the complete DM density $\Omega_{DM}\,h^2$ given in (\ref{relic}).
The color code is as follows: red points (dark gray) produce a relic density
above the $3\sigma$ limit given in eq.~(\ref{relic}); blue points (black) produce a relic density within
the $3\sigma$ region;
green points (light gray) produce a relic density below the $3\sigma$ limit. Regarding the points that satisfy
the relic density we see three clear regions \cite{Honorez:2010re, LopezHonorez:2010tb}, one for low $m_H$ ($3<m_H<50$ GeV),
another one for medium $m_H$ ($60<m_H<100$ GeV) and finally one for high values of $m_H$
($m_H>550$ GeV). The explanation for the gap is related to annihilation processes and it
will be given later. At $m_H<3$ GeV the IHDM can no longer be compatible with
vacuum existence and stability \cite{Goudelis:2013uca, Cirelli:2005uq}.

%
\begin{figure}[ht]
\includegraphics[height=6.5cm]{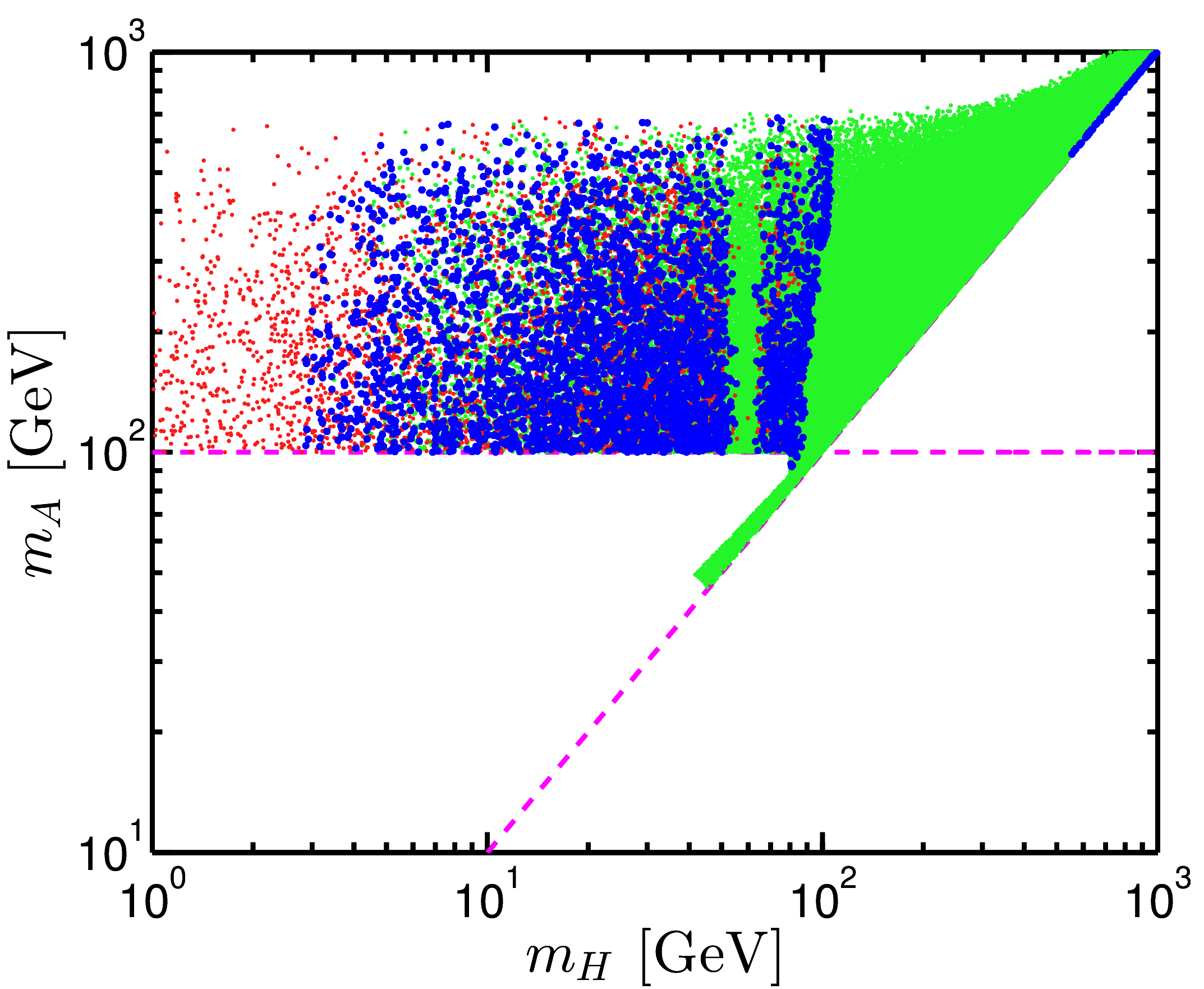}\hfill
\includegraphics[height=6.5cm]{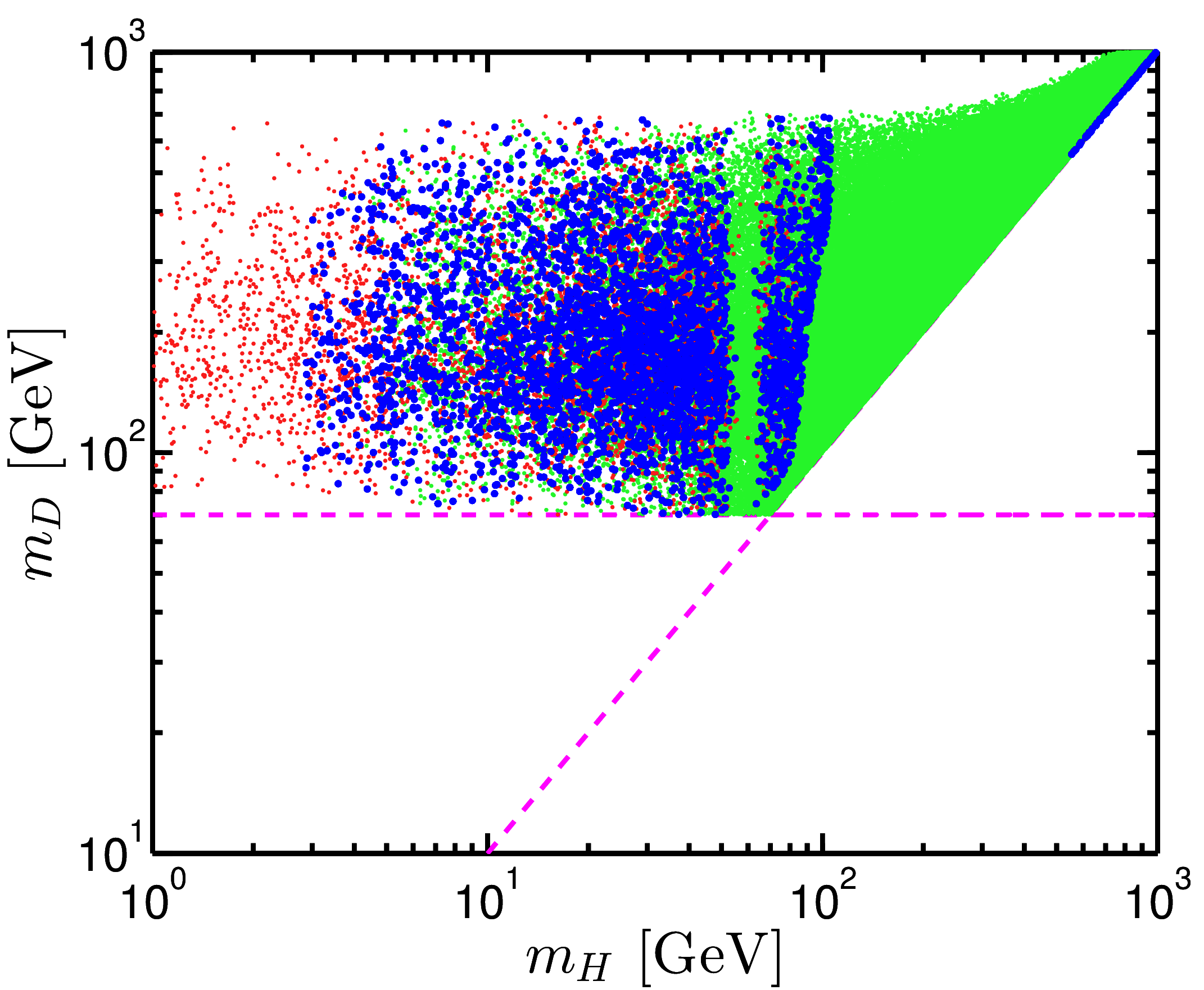}
\caption{\it Inert Higgs masses $m_A$ (left panel) and $m_D$ (right panel) as a function of the DM
candidate mass $m_H$, for the same random scan of IHDM parameter space.
The horizontal dotted line is due to LEP constraints and the diagonal dotted line is due to the $m_H$-DM  condition.}
\label{FigB}
\end{figure}
%
In Fig.~\ref{FigB} we have for the same scan and color code the mass of the heavy pseudoscalar $m_A$ (left) and the
mass of the charged Higgs $m_D$ (right) as a function of the mass of the DM candidate $m_H$. Due to dedicated
pre-LHC collider searches, a bound that captures most of the features is $m_A>100$ GeV and $m_D>70$ GeV. This
is so with the exception of a small strip for $m_A<100$ GeV seen in the left panel of Fig.~\ref{FigB}, where
due to the proximity of the $H$ and $A$ masses the search looses sensitivity. In this figure the gap in values
of $m_H$ when the relic density is imposed is apparent. Notice that the density of solutions is larger when
the masses for $H$, $A$, and $D$ are close to each other, and that this feature is more pronounced when the
masses of these particles are near the TeV scale (due to the logarithmic scale). Finally we notice that
$\Omega_{DM}\,h^2$ is more sensible
to  the parameters $\lambda_{345}$ and $m_H$ (Fig.~\ref{FigA}) than the masses of the other inert particles
(Fig.~\ref{FigB}).

\section{Collider Physics}
\label{secCollider}

As we mentioned before, the Higgs boson discovered at CERN in 2012 is the SM-like Higgs boson $h$ of our model
from the non-inert Higgs doublet field $\Phi_S$. This particle $h$ couples to the charged Higgs pair ($D^\pm$),
which contributes to the diphoton decay width\footnote{In this scenario, the IHDM cannot
reproduce the reported excess of diphoton events by ATLAS \cite{atlas_exceso} and CMS \cite{cms_exceso}
Collaborations in their Run-II $13$ TeV analyses, because the $\mathbb{Z}_2$ symmetry
prevent the extra Higgs bosons of the model from decaying into two photons.}
$\Gamma(h\rightarrow\gamma\gamma)$ \cite{HHG}. For the same reason $D^\pm$
also contribute to $\Gamma(h\rightarrow Z\gamma)$.

It is convenient to work with the parameter \cite{Arhrib:2012ia, Goudelis:2013uca, Krawczyk:2013jta}
\begin{equation}
R_{\gamma\gamma} =
\dfrac{B(h\to\gamma\gamma)^\mathit{IHDM}}{B(h\to\gamma\gamma)^\mathit{SM}}\quad.
\label{Rgg}
\end{equation}
The value we use for the SM is $\Gamma(h_{SM}\rightarrow\gamma\gamma)=4.1$ MeV \cite{Djouadi:1997yw}.
ATLAS \cite{atlas:ss} and CMS \cite{cms:ss} collaborations have studied this decay mode, and if we
combined both results \cite{stat} obtain $R_{\gamma\gamma}^\mathit{exp}=1.14\pm0.18$.

%
\begin{figure}[ht]
\includegraphics[height=6.5cm]{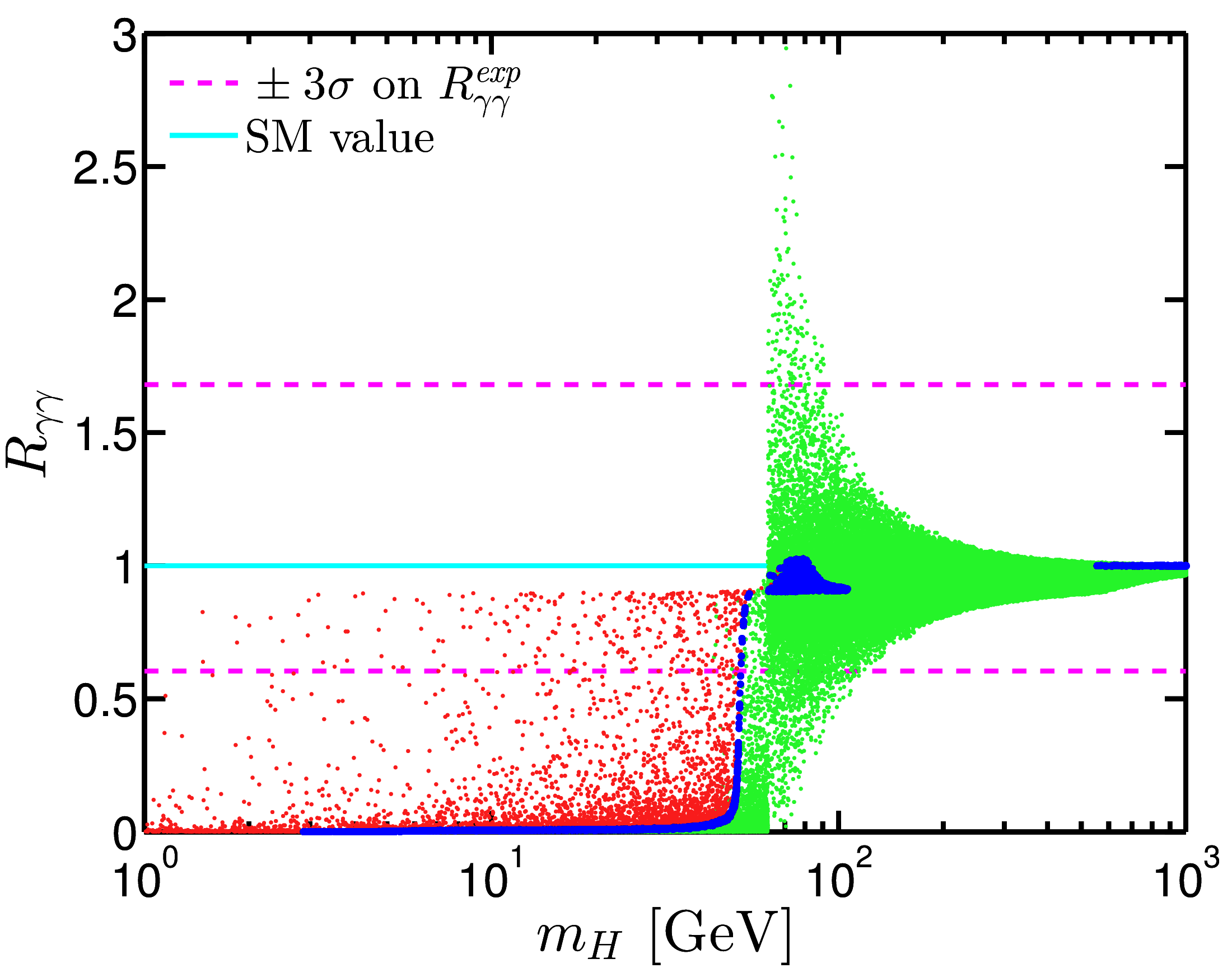}\hfill
\includegraphics[height=6.5cm]{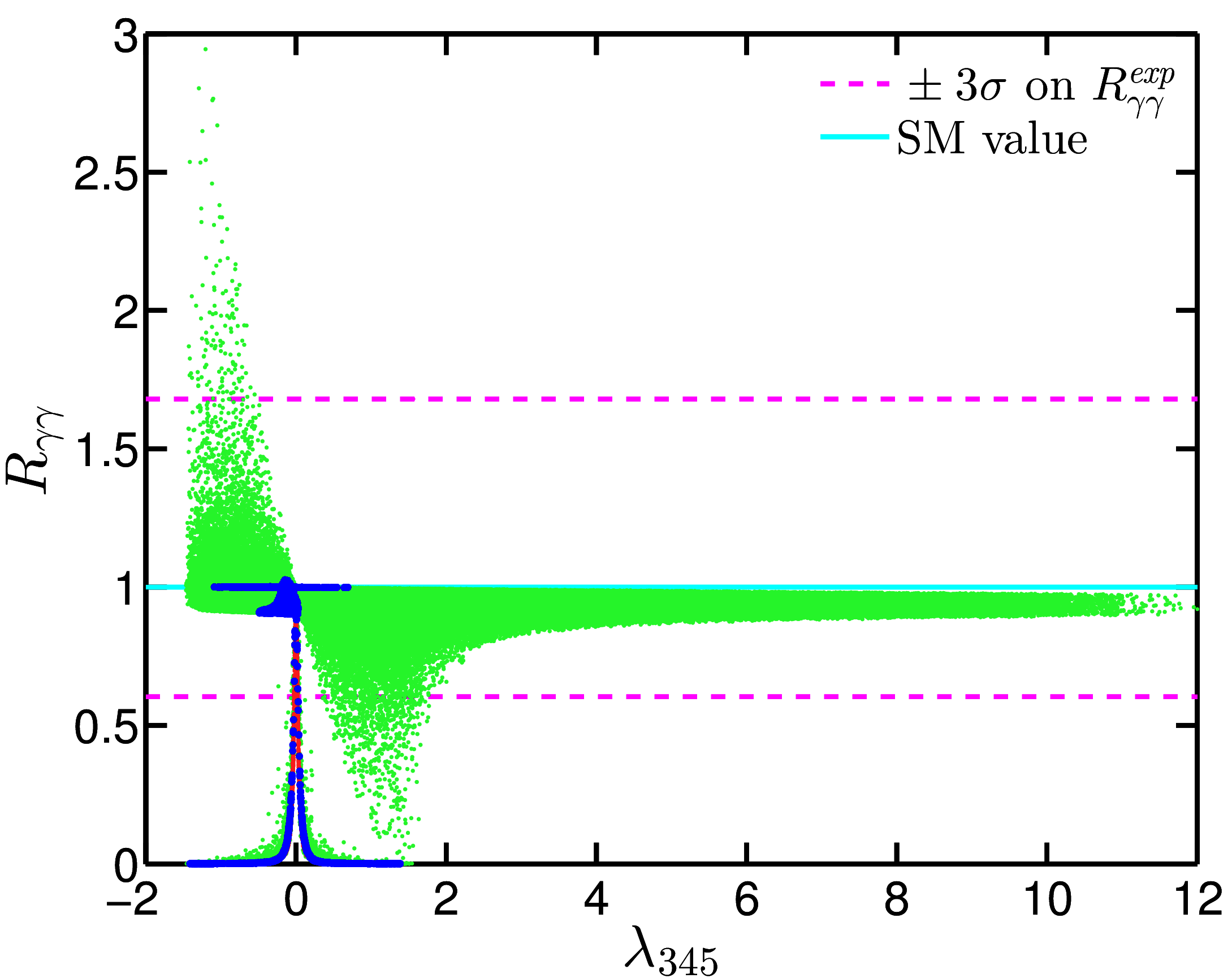}
\caption{\it $R_{\gamma\gamma}$ parameter as a function of the DM candidate mass $m_H$ (left panel)
and the $\lambda_{345}$ coupling (right panel), for the same random scan of IHDM parameter space.}
\label{FigC}
\end{figure}
%
In Fig.~\ref{FigC} we have the parameter $R_{\gamma\gamma}$ as a function of the DM candidate mass $m_H$
(left panel) and as a function of the coupling $\lambda_{345}$ (right panel). The points in parameter
space that produce a correct relic density can be divided in three groups. In the case of very light
masses for the DM candidate ($3<m_H<50$ GeV approximately) the decay mode $h\rightarrow HH$ is open,
and $R_{\gamma\gamma}$ is close to zero ruling those masses out \cite{Ilnicka:2015jba}. In the intermediate mass case,
approximately between 60 and 100 GeV, there is a region with acceptable solutions characterized by
$R_{\gamma\gamma}\approx 1$. This region is characterized by increasingly heavier values for $A$ and
$D^\pm$. In the large mass region ($m_H>550$ GeV approximately), the charged Higgs $D^\pm$ gives a
negligible contribution to the decay $h\rightarrow\gamma\gamma$ such that $R_{\gamma\gamma}$ is close
to unity. Interestingly, the right panel of Fig.~\ref{FigC} shows that perturbative values ($|\lambda_{345}|<1$)
are preferred.

\begin{figure}[ht]
\includegraphics[height=6.5cm]{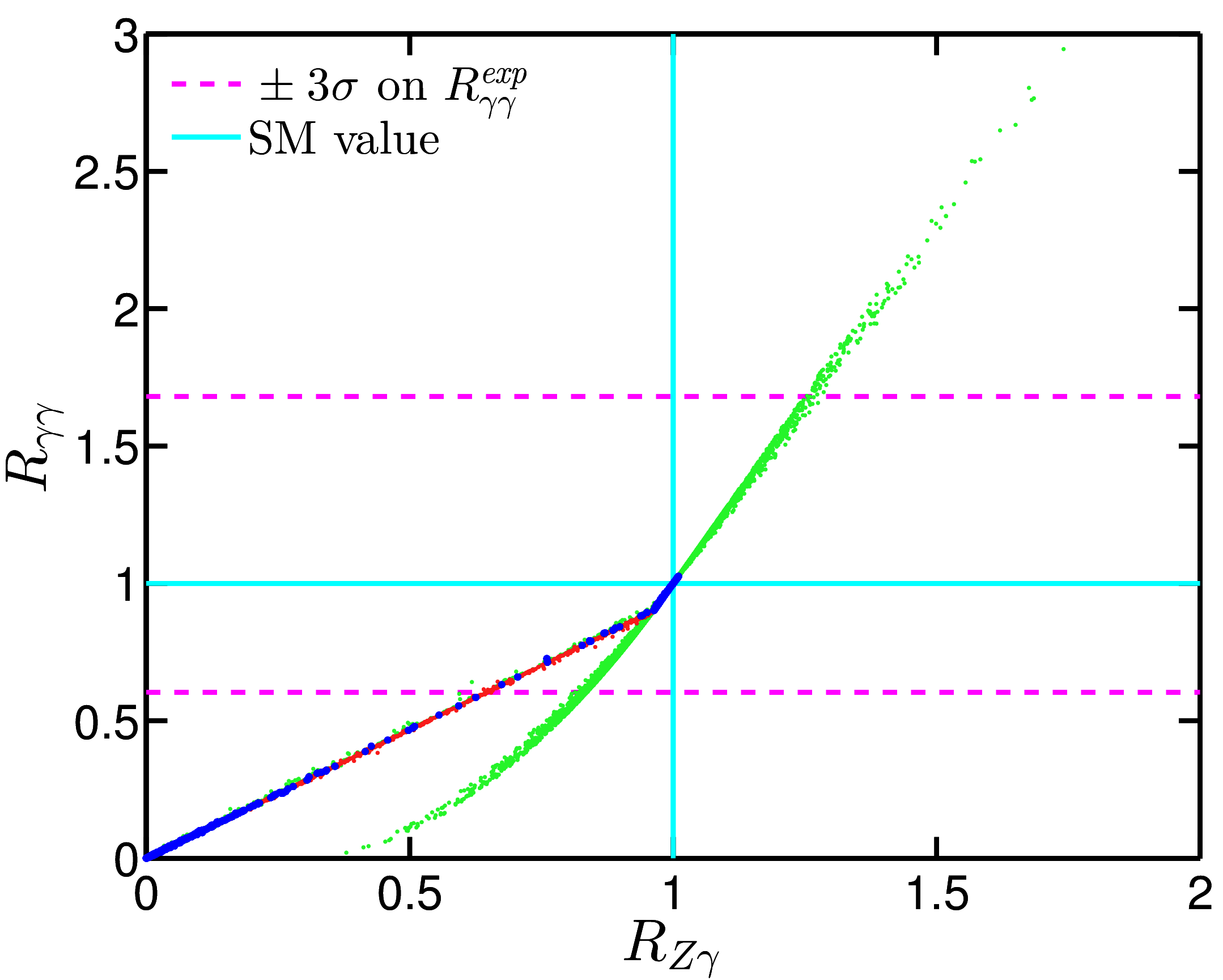}
\caption{\it Relation between $R_{\gamma\gamma}$ and $R_{Z\gamma}$ for the same random scan of IHDM
parameter space.}
\label{FigD}
\end{figure}

In addition, if the inert particles are light enough, there are
two other two-body decays which are:
\begin{equation}
\begin{array}{rcl}
\espaBig \Gamma(h\to HH) &=& \dfrac{v^2\,{\lambda_{345}}^2}{32\pi\,m_h}\sqrt{1-\dfrac{4{m_H}^2}{{m_h}^2}}\quad, \\
\espaBig \Gamma(h\to AA) &=& \dfrac{\Big({m_A}^2-{m_H}^2+{\lambda_{345}}\,v^2/2\Big)^2}{8\pi
v^2\,m_h}\sqrt{1-\dfrac{4{m_A}^2}{{m_h}^2}}\quad.
\end{array}
\end{equation}
There is no phase space for a two-body decay $h\rightarrow D^\pm D^\mp$.
It is possible to define the parameter $R_{Z\gamma}$, in analogy to $R_{\gamma\gamma}$ defined in
eq.~(\ref{Rgg}).
It is interesting that even though the decay $h\rightarrow Z\gamma$ is not well measured,
it still can give additional insight to the model \cite{atlas_rzg,cms_rzg}.
As Fig.~\ref{FigD} shows, there appears a very narrow correlation between $R_{\gamma \gamma}$ and $R_{Z\gamma}$,
which is a common feature for $R_{\gamma \gamma}$ v/s $R_{Z\gamma}$ plots \cite{Chen:2013vi, Fortes:2014dia, Krawczyk:2013pea}.
%
%
The bisector branch only
contains points that satisfy $m_H<m_h/2$ (inert invisible decay channel open). Points of parameter space
which satisfy the relic density and have low DM candidate mass are ruled out, because they produce a too small
value for $R_{\gamma\gamma}$.
By analyzing the characteristics of the those data points one finds that the
larger branch includes only points with $m_H>m_h/2$, and the ones that also satisfy relic density are
close to $R_{\gamma\gamma}=R_{Z\gamma}=1$, as it was mentioned before.
The two branches seen in the $R_{\gamma\gamma}$ v/s $R_{Z\gamma}$ relation also appear in the non-normalized
$B(h\rightarrow\gamma\gamma)$ v/s $B(h\rightarrow Z\gamma)$ relation (not shown). But it is reduced only to the long
(green) branch in the $\Gamma(h\rightarrow\gamma\gamma)$ v/s $\Gamma(h\rightarrow Z\gamma)$ relation.
Further one sees that when the $h\rightarrow HH$ decay channel is closed, the $D^\pm$ loop transforms into the long
branch the otherwise SM dot (the intersection point between the two branches). If the $h\rightarrow HH$
decay channel is open, the second branch appears because the $h\rightarrow HH$ channel tends to dominate \cite{Swiezewska:2012eh}.

We are also interested in the invisible decay of the SM-like Higgs boson. If the DM candidate mass
satisfy $m_H<m_h/2$, the two-body decay channel $h\rightarrow HH$ is open, which is invisible for the
LHC detectors, and shows only as missing momentum. There are measurements for the invisible decay of the
SM-like Higgs from the LHC experiments. Taking a simple average of the upper bounds to the invisible
decay rate from ATLAS \cite{atlas:inv} and CMS \cite{cms:inv}, gives $B(h\rightarrow \rm{inv})<0.43$
for the SM-like Higgs boson.
%
\begin{figure}[ht]
\includegraphics[height=6.5cm]{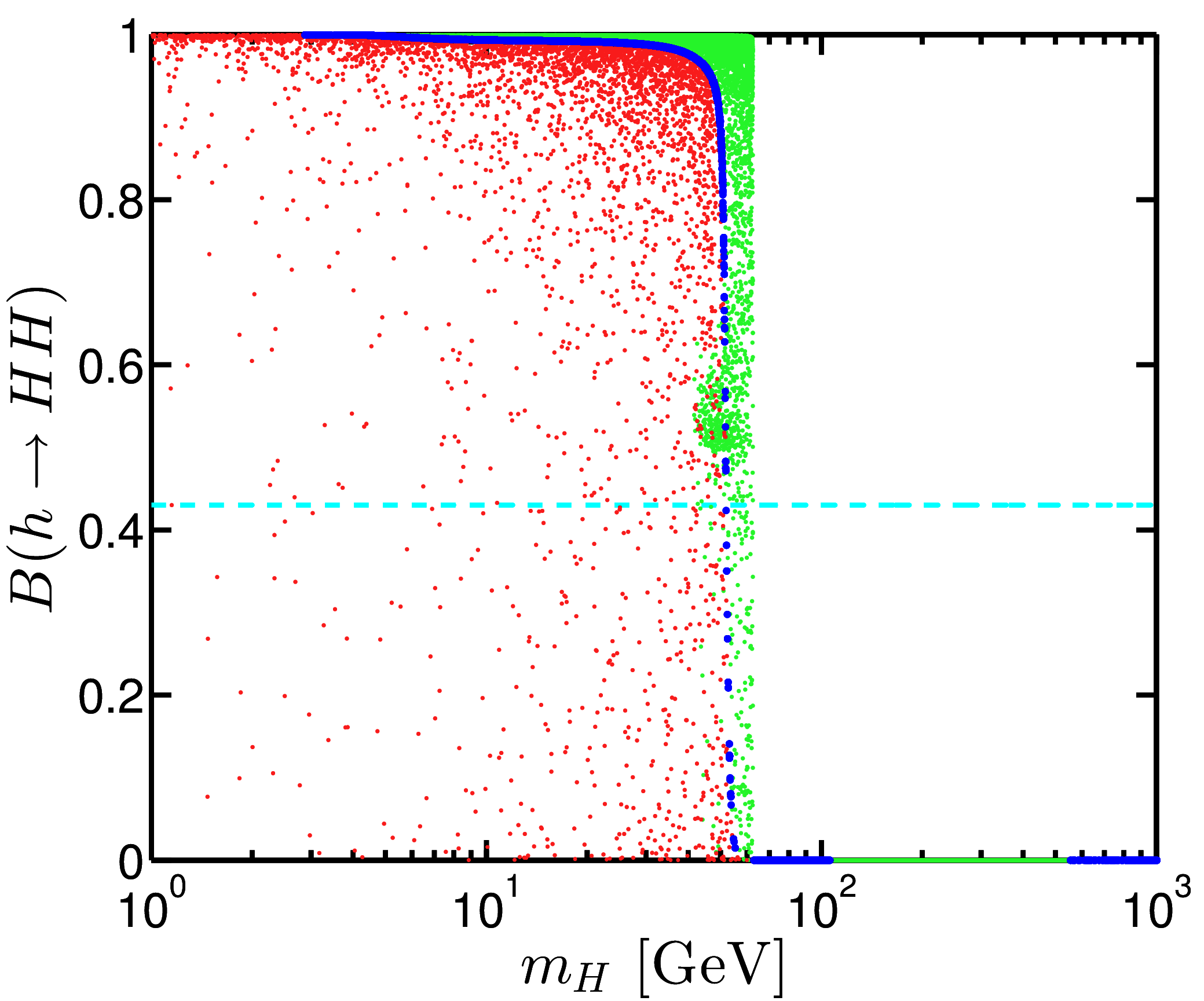}\hfill
\includegraphics[height=6.5cm]{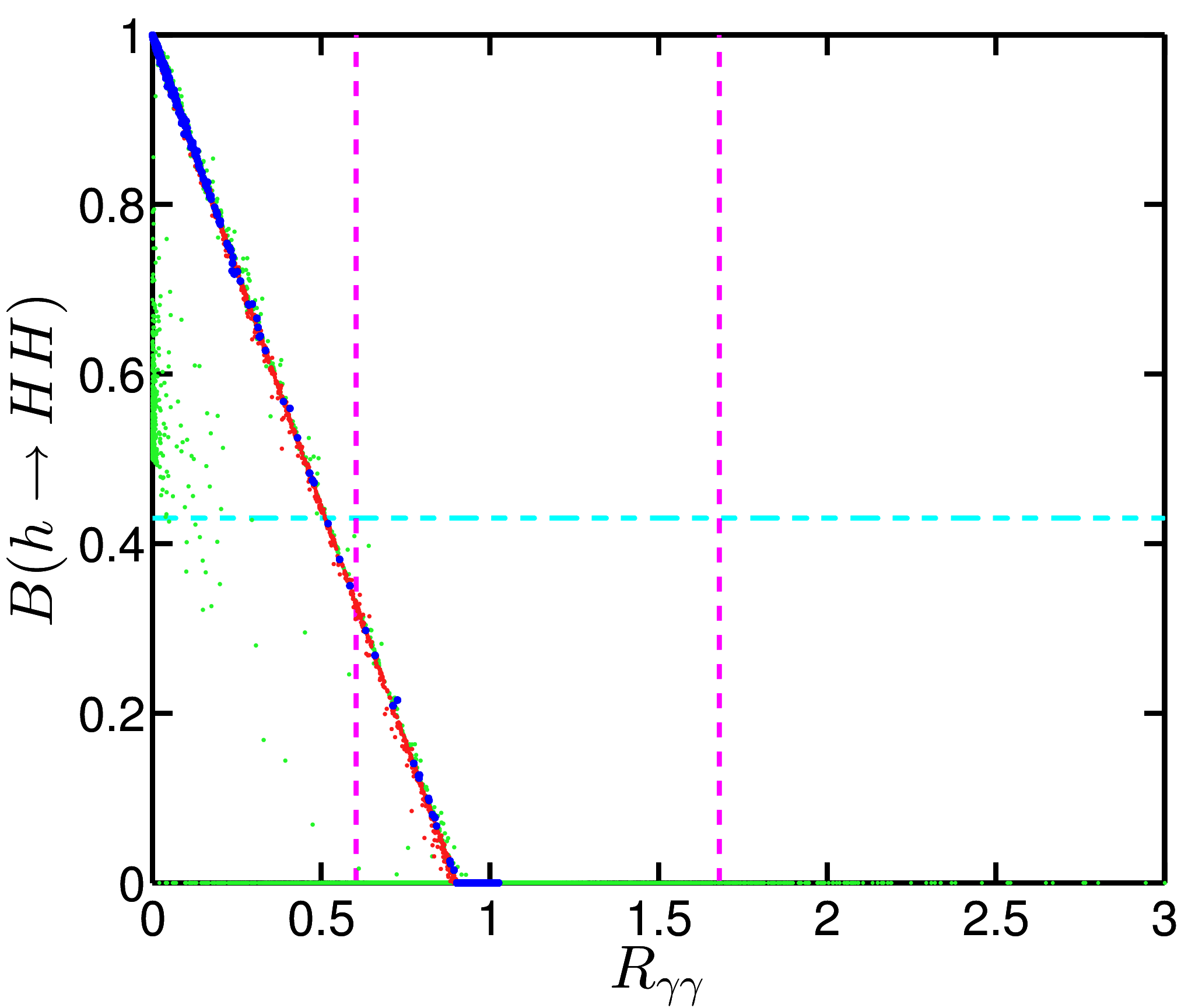}
\caption{\it Branching ratio for the invisible decay of the SM-like Higgs boson as a function of the
DM candidate mass $m_H$ (left panel) and as a function of the parameter $R_{\gamma\gamma}$ (right panel),
for the same random scan of IHDM parameter space.
Please note that there is a number of points on the $m_H$ and $R_{\gamma \gamma}$ axes,
with $B(h\rightarrow HH)=0$.
}
\label{FigE}
\end{figure}
%
In Fig.~\ref{FigE} we show the branching ratio for the invisible decay of the SM-like Higgs boson
$B(h\rightarrow HH)$, as a function of the mass of the DM candidate $m_H$ (left panel) and as a function
of the parameter $R_{\gamma\gamma}$ (right panel). In the left panel we also have a horizontal line that
shows the upper bound for $B(h\rightarrow HH)$ mentioned above. The threshold $2m_H=m_h$ appears clearly
in the left panel. Most of the points with correct relic density satisfying $m_H<60$ GeV are ruled out
because they produce a too large invisible branching ratio for $h$. On the contrary, most of the points
with higher DM candidate mass are fine because they produce an invisible  branching ratio equal to
zero. In the right panel, where we have as dashed lines the bounds from LHC experiments,
we see a very strong relation between $B(h\rightarrow HH)$ and $R_{\gamma\gamma}$. The
points that satisfy relic density with a low mass for the DM candidate are simultaneously excluded
from $B(h\rightarrow HH)$ and from $R_{\gamma\gamma}$. The rest of the points satisfy the bounds.

%
\begin{figure}[ht]
\includegraphics[height=5.6cm]{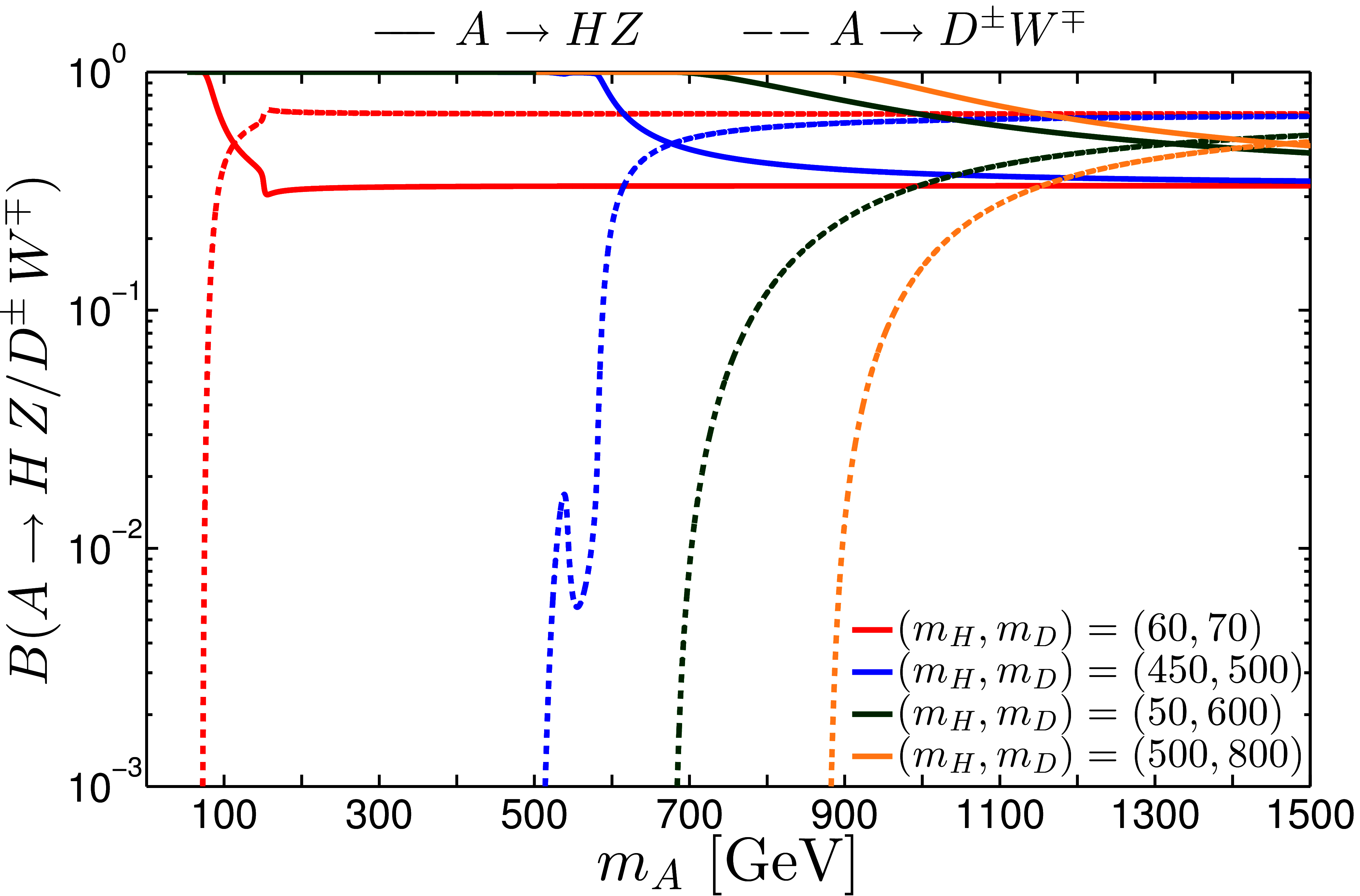}\hfill
\includegraphics[height=5.6cm]{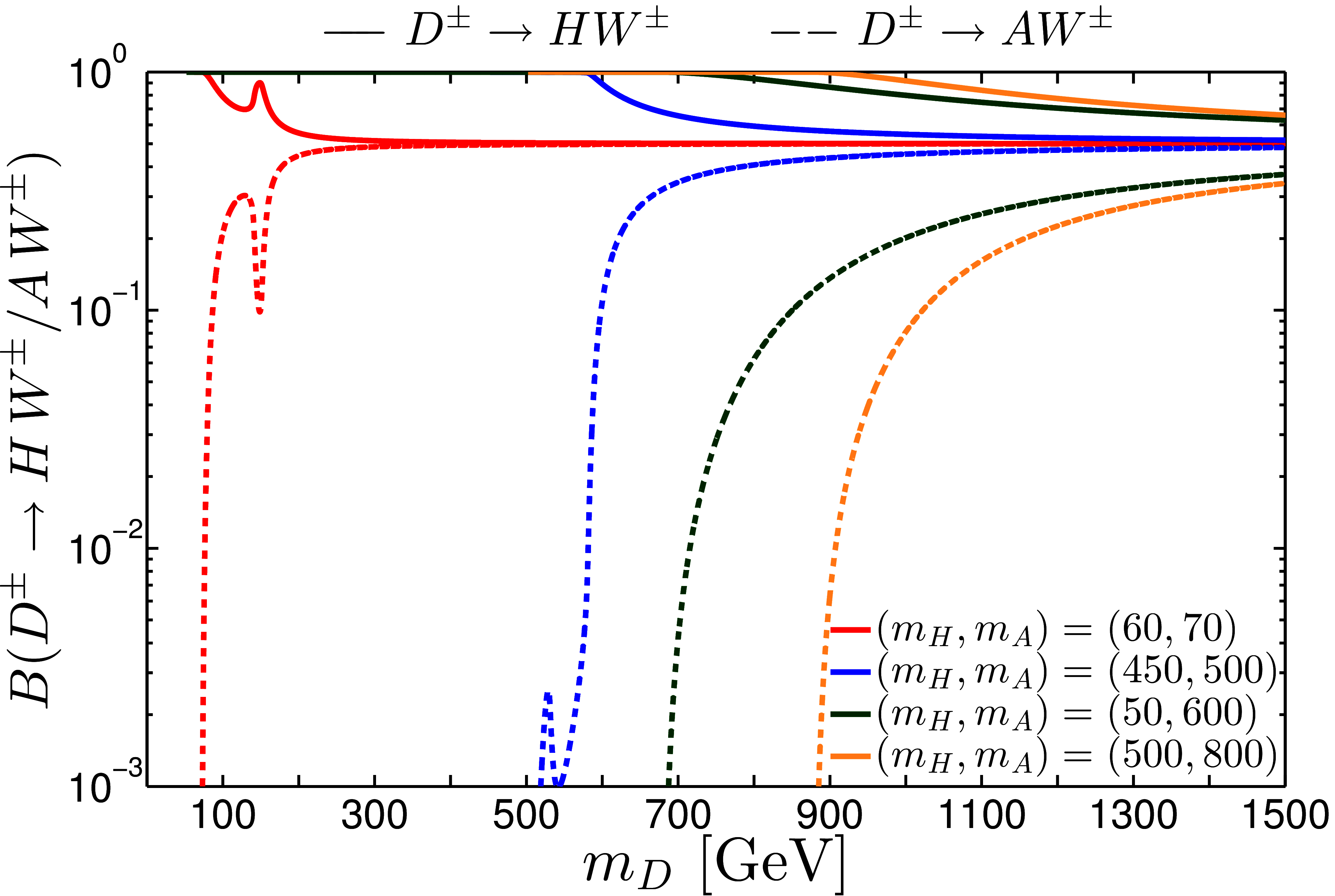}
\caption{\it Branching ratios for the inert Higgs bosons $A$ (left panel) and $D^\pm$ (right panel)
as a function of their corresponding mass.}
\label{FigF}
\end{figure}
%
To finalize this section we discuss the branching ratios for the two observable inert Higgs bosons
$A$ and $D^\pm$. In Fig.~\ref{FigF} we have the branching ratios for the pseudoscalar Higgs boson
(left panel) as a function of its mass,
and the branching ratios for the charged Higgs boson
(right panel) as a function of its mass. In the left panel we show the decays of the inert pseudoscalar Higgs $A$,
which are $A\rightarrow HZ$ (solid line) and $A\rightarrow D^\pm W^\mp$ (dashed line).
As it can be seen from the Feynman rules,
 the only unknown parameters the branching ratios depend on are the masses $m_H$, $m_A$, and $m_D$.
 Therefore,
four scenarios are considered: (i) $(m_H,m_D)=(60,70)$, (ii) $(450,500)$, (iii) $(50,600)$, and (iv)
$(500,800)$ GeV. The gauge boson can be off-shell, although we consider the inert Higgs bosons always on-shell.
The oscillation near the threshold is due to different increasing rates for the decay rates when the
gauge boson is off-shell. The branching ratio $B(A\rightarrow HZ)$ is always large (because $m_H<m_D$)
while $B(A\rightarrow D^\pm W^\mp)$ can be low near thresholds. There is a crossing point where
$B(A\rightarrow HZ)=B(A\rightarrow D^\pm W^\mp)$. In the right panel we show the decays of the charged
Higgs $D^\pm$. Analogous scenarios are considered, but replacing $m_D$ by $m_A$. In solid line we have
the branching ratio for the decay $D^\pm\rightarrow HW^\pm$ and in dash we have
$D^\pm\rightarrow AW^\pm$. In the case of $D^\pm$, there is no crossing point, thus
$B(D^\pm\rightarrow HW^\pm)$ is always larger than $B(D^\pm\rightarrow AW^\pm)$. We remind the reader
that the presence of a Higgs boson $H$ in a final state is seen as missing momentum at the LHC.

\section{Cosmology and Dark Matter}
\label{secDM}

The existence of Dark Matter seems to be well established now \cite{Agashe:2014kda}. There are several
candidates for DM, among them the previously mentioned WIMPs.
A good particle candidate for DM
must be neutral and stable (or quasi-stable). The $\mathbb{Z}_2$ discrete symmetry in the model studied
in this article ensures that the lightest of the inert Higgs bosons is stable. Observation implies it is either
$H$ or $A$ (or in a fine tuned scenario both). In this article we study the former case. An important restriction this candidate must
satisfy is that its mass density must agree with experimental observations. We calculate the relic
density of our DM candidate using \verb"micrOMEGAs" software \cite{micromegas}. To better understand the
results on the relic density, we calculate also the thermal averaged annihilation cross section times
the relative velocity $\langle\sigma v\rangle$, or annihilation cross section for short.

%
\begin{figure}[ht]
\includegraphics[height=6.5cm]{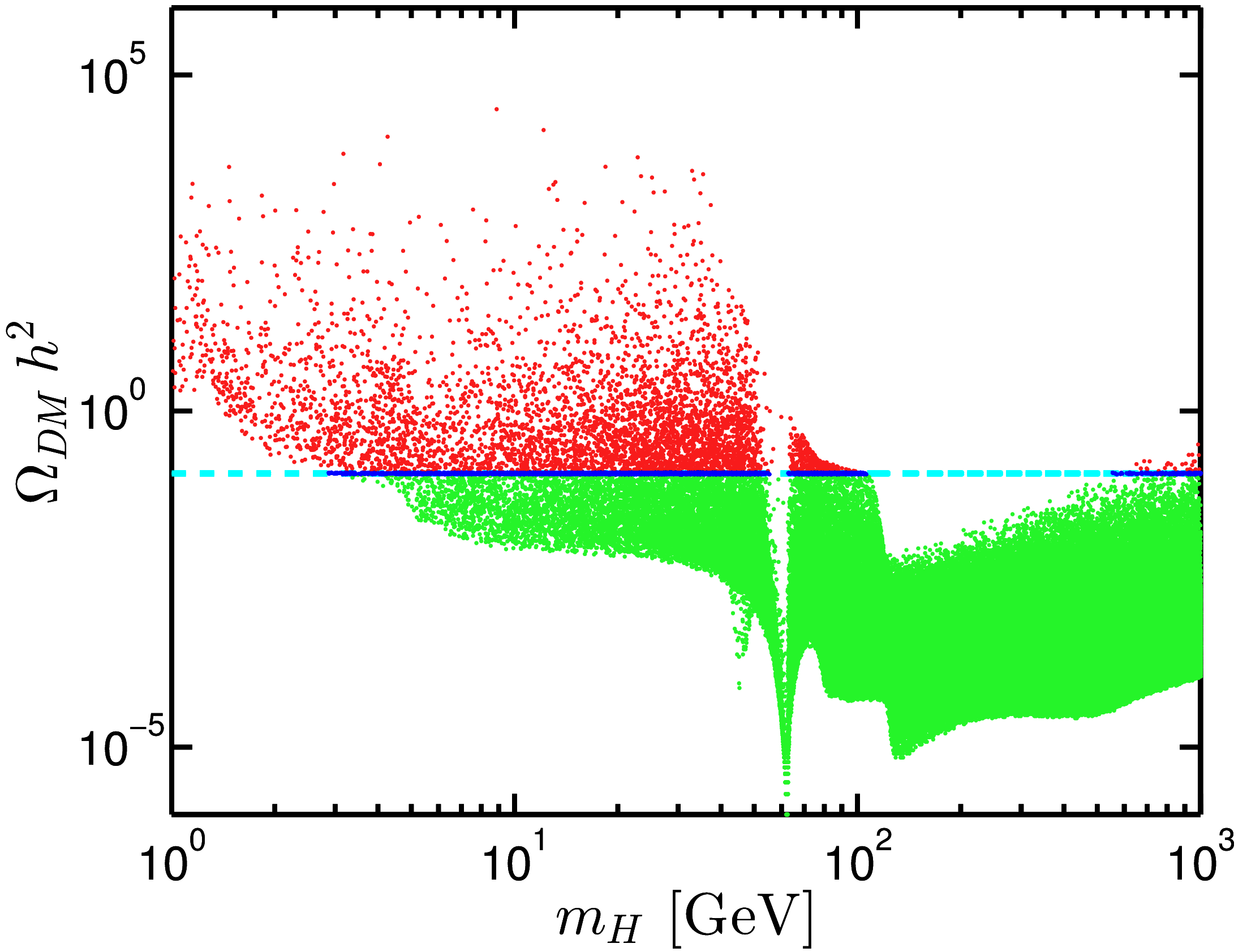}\hfill
\includegraphics[height=6.65cm]{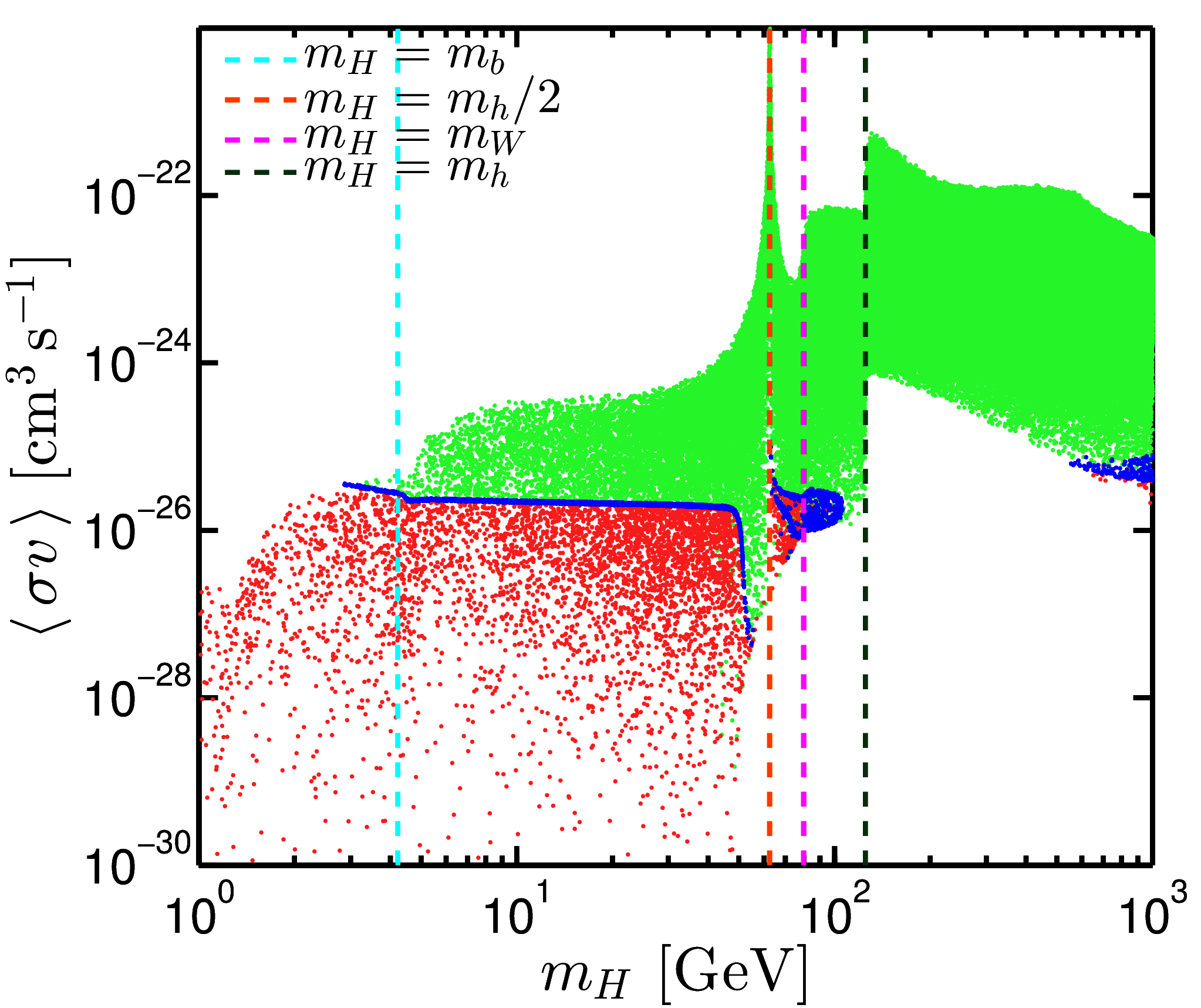}
\caption{\it Relic density for $H$ as a function of its mass $m_H$ (left panel), and annihilation
cross section for $H$ also as a function of its mass $m_H$, for the same random scan of IHDM
parameter space.}
\label{FigG}
\end{figure}
%
In Fig.~\ref{FigG} we plot the $H$ relic density as a function of the DM candidate mass $m_H$ (left
panel), and the annihilation cross section also as a function of the DM candidate mass $m_H$ (right panel),
calculated with \cite{micromegas}.
For the relic density case, we also show, as an horizontal dashed line, the experimentally measured value for
$\Omega_{DM} h^2$ as given in eq.~(\ref{relic}).
The scan shows a large distribution with differences
that can reach more than 10 orders of magnitude. For this reason most of the points in the scan are
ruled out if one demands that $H$ is actually the only WIMP responsible for the observed
DM signatures.
There are two mass gaps that divide the mass region in three: low mass ($3<m_H<50$ GeV approximately), medium mass
($60<m_H<100$ GeV approximately), and high mass ($550<m_H$ approximately).
The origin of these
mass gaps is better understood with the aid of the right frame. In the right frame of Fig.~\ref{FigG}
we show the annihilation cross section as a function of the DM candidate mass $m_H$, with vertical
lines denoting different thresholds. The first gap is near the threshold $m_H\approx m_h/2$ where the
annihilation channel $HH\rightarrow h$ becomes very efficient due to the on-shellness of the SM-like
Higgs $h$. The second gap starts at the thresholds $m_H\approx m_W$ and $m_H\approx m_Z$, where $HH\rightarrow WW(ZZ)$
become available, and continue later with the threshold $m_H\approx m_h$ where the channel $HH\rightarrow hh$
opens up. The annihilation channel $HH\rightarrow t\overline{t}$ also helps. All these new annihilation
channels make the DM annihilation very efficient, and it is not possible to obtain a relic density
according to observations. On the other hand, for larger $m_H$, it is possible to get a correct relic density if the difference between the three inert scalar masses is not so large and $\lambda_{345}$ remains small enough \cite{Hambye:2009pw} (see Fig.~\ref{FigA}).

We finally study the direct detection prospects of our DM candidate. We do that through the tree-level spin-independent
DM-nucleon interaction cross section \cite{Djouadi:2011aa}, which applied to our case is,
\begin{equation}
{\sigma^\textit{SI}}_\textit{DM-N}=\dfrac{{\lambda_{345}}^2}{(4\pi {m_h}^4)}\dfrac{{m_N}^4\,{f_N}^2}{(m_H+m_N)^2}\quad.
\label{scattering}
\end{equation}

Here $m_h$ is the mass of the SM-like Higgs boson, $m_H$ is the mass of the DM candidate, $m_N$ is
the nucleon mass, taken here to be $m_N=0.939$ GeV as the average of the proton and neutron masses,
$\lambda_{345}$ is the combined coupling
defined in eq.~(\ref{lambdaT}), and $f_N$ is a form factor that depends on hadronic matrix elements
\cite{Kanemura:2010sh, Alarcon:2012nr}.

%
\begin{figure}[ht]
\includegraphics[height=6.5cm]{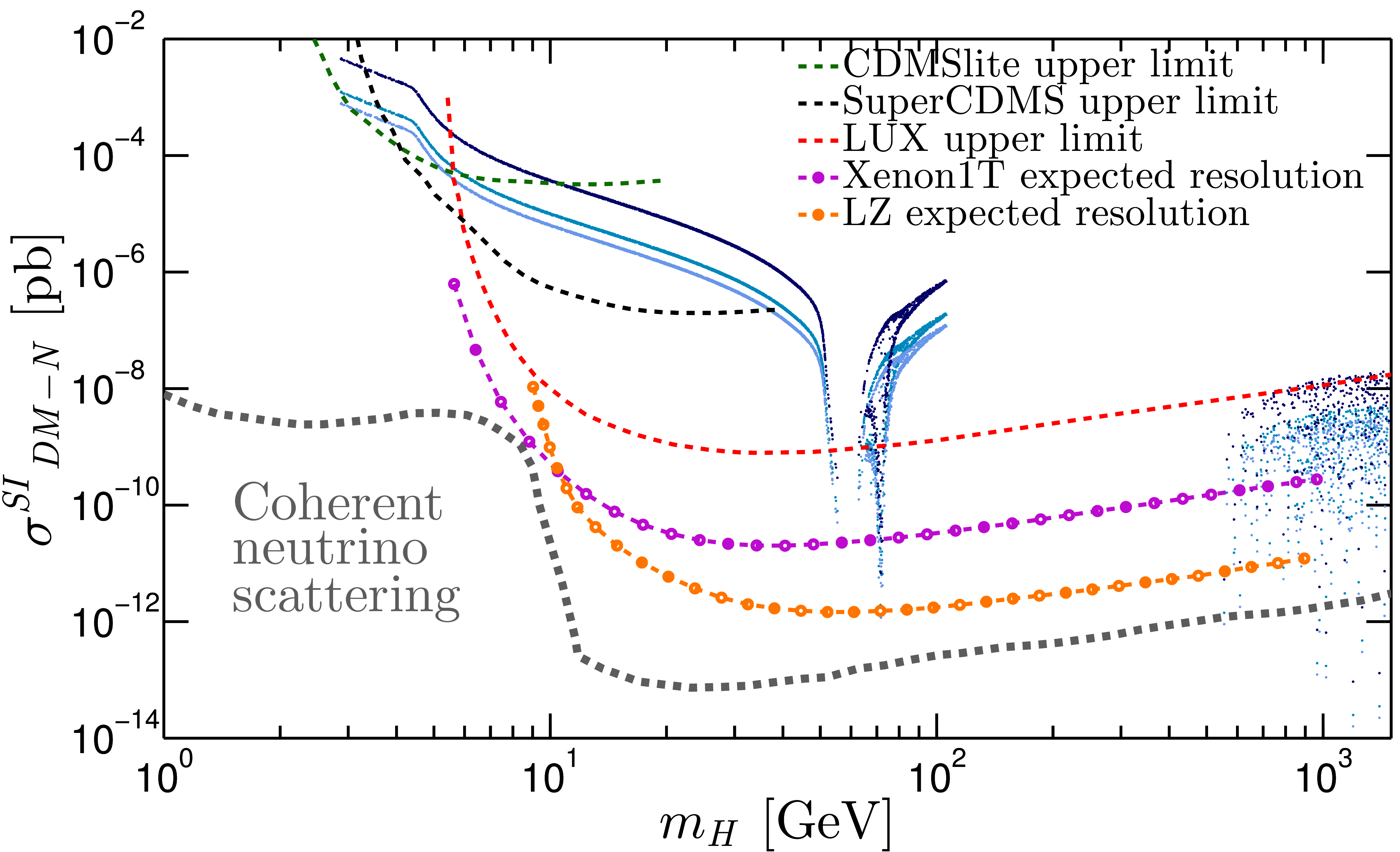}
\caption{\it Spin-independent DM-nucleon cross section as a function of the DM candidate mass, for
points of the IHDM with DM relic density consistent with observation. Three scenarios are considered
with different values for the form factor $f_N$. Lower bounds for past experiments and prospects of
measurements for future experiments are shown.}
\label{FigH}
\end{figure}
%

In Fig.~\ref{FigH} we show the DM-nucleon cross section as a function of the DM candidate mass
for a correct value of $\Omega_{DM}\,h^2$. We
consider three values for $f_N$: a central value (0.326) from a lattice calculation \cite{Young:2009zb},
and extreme values (0.260 and 0.629) from the MILC collaboration \cite{Toussaint:2009pz}.
Lower bounds for past experiments and prospects of measurements for future experiments are also
shown \cite{lux2013, cdms_lite, super_cdms, dmtools}.
Notice that the dispersion of points for high $m_H$ can be understood
from analogous dispersion seen in Fig.~\ref{FigA}, and the same situation
occurs with the line-like distribution for light $m_H$.
We also show the coherent neutrino scattering upper limit \cite{Billard:2013qya}.
This curve represents the threshold below which the detector sensitivity is such that not only the
possible DM scattering effects can be observed,
but also the indistinguishable scattering effects associated with neutrinos.
Thus, this indicates a region where the neutrino background becomes dominant and
and little information can be obtained on DM effects.
Current
direct detection of DM exclude all the low DM mass points, and most of the medium DM mass points. Allowed
are a narrow region near 60 GeV, and all the high mass region.
Note, that the absence of points in the range of $\sim 100- 550$~GeV in this plot
is due to the fact that the plotted points are only those that give the right dark matter density (``blue points'').
Future experiments will be able to test large parts of these two regions,
but will not be able to rule them out entirely if there is no signal.

\section{Conclusions}
\label{secConc}

In this article the Inert Higgs Doublet Model is studied, with the inert Higgs boson $H$ as a DM
candidate, using the latest results for DM relic density, annihilation cross section, and collider searches.
As a summary we highlight,
\begin{itemize}
\item
The branching ratios for the charged Higgs $D^\pm$ and for the pseudoscalar Higgs
$A$ are studied and shown in  Fig.~\ref{FigF}. The $\mathbb{Z}_2$ symmetry strongly reduces the number of different decay channels. Considering the Higgs
boson on-shell and allowing the gauge boson to be off-shell (a different choice would produce different decay
channels, but with smaller branching ratios), we find that $B(D^\pm\to HW^\pm) > B(D^\pm\to AW^\pm)$ as opposed
to the $A$ decays, where there is a crossing point not far from the threshold. For this reason, in collider
searches we recommend to look for a signal for a $D^\pm$: two jets (consistent with a $W$) and missing energy
(from the DM candidate $H$).
\item

Three distinct $H$ mass regions are found that produced the correct relic density ({\ie} $H$ is
the only source for DM) as it can be seen in Fig.~\ref{FigG}. The low mass region (between 3 and 50 GeV approximately) is already ruled out because
it produces a too small value for $R_{\gamma\gamma}$ (Fig.~\ref{FigC}), because it produces a too high value
for $B(h\rightarrow HH)$ (Fig.~\ref{FigE}), and because of direct DM searches (Fig.~\ref{FigH}). The
intermediate mass region (between 60 and 100 GeV approximately)
and the high mass region (heavier than 550 GeV approximately) are allowed.
\item
In Fig.~\ref{FigH} we study the DM candidate direct detection. The low mass region is also ruled out by present
experiments. In addition, future experiments will probe intermediate and high mass regions. Nevertheless, in absence
of signals, it will not be possible to rule out these two regions. Notice the proximity of the region where
the coherent neutrino scattering is an irreducible background.
\end{itemize}
With this one sees that the Inert Higgs Doublet Model gives a still viable DM candidate,
which will most likely be tested by direct DM detection experiments.

\begin{acknowledgments}
We thank conversations with Drs.~Germ\'an G\'omez, Nicol\'as Viaux, and Edson Carqu\'{\i}n.
M.A.D. was partly supported by Fondecyt 1141190.
B.K. was partly supported by Fondecyt 1120360.
S.U-Q. was partly supported by postgraduate CONICYT grant.
The work of all of us was also partly financed by CONICYT grant ANILLO ACT-1102.
\end{acknowledgments}


\end{document}